\documentclass[aip,aps,preprint,showpacs]{revtex4}
\usepackage{amssymb}

\usepackage{graphicx}

\usepackage[T1]{fontenc}

\usepackage[american]{babel}

\begin{document}

\title{Growth of Ca$_2$MnO$_4$ Ruddlesden-Popper structured thin films using Combinatorial Substrate Epitaxy}

\author{M.~Lacotte}
\affiliation{Laboratoire CRISMAT, CNRS UMR 6508, ENSICAEN, Universit$\acute{e}$ de Basse-Normandie, 6 Bd Mar$\acute{e}$chal Juin, F-14050 Caen Cedex 4, France.}
\author{A.~David}
\affiliation{Laboratoire CRISMAT, CNRS UMR 6508, ENSICAEN, Universit$\acute{e}$ de Basse-Normandie, 6 Bd Mar$\acute{e}$chal Juin, F-14050 Caen Cedex 4, France.}
\author{D.~Pravarthana}
\affiliation{Laboratoire CRISMAT, CNRS UMR 6508, ENSICAEN, Universit$\acute{e}$ de Basse-Normandie, 6 Bd Mar$\acute{e}$chal Juin, F-14050 Caen Cedex 4, France.}
\author{C.~Grygiel}
\affiliation{Laboratoire CIMAP, CEA, CNRS UMR 6252, ENSICAEN, Universit$\acute{e}$ de Basse-Normandie, 6 Bd Mar$\acute{e}$chal Juin, F-14050 Caen Cedex 4, France}
\author{G.S.~Rohrer and P.A.~Salvador}
\affiliation{Department of Materials Science and Engineering, Carnegie Mellon University, 5000 Forbes Ave., Pittsburgh, Pennsylvania 15213.}
\author{M.~Velazquez}
\affiliation{CNRS, Universit$\acute{e}$ de Bordeaux, ICMCB, UPR 9048, F-33600 Pessac, France}
\author{R.~de Kloe}
\affiliation{Application laboratory, Tilburg, The Netherland}
\author{W.~Prellier}\thanks{wilfrid.prellier@ensicaen.fr} 
\affiliation{Laboratoire CRISMAT, CNRS UMR 6508, ENSICAEN, Universit$\acute{e}$ de Basse-Normandie, 6 Bd Mar$\acute{e}$chal Juin, F-14050 Caen Cedex 4.}

\date{\today}

\begin{abstract}
The local epitaxial growth of pulsed laser deposited Ca$_2$MnO$_4$ films on polycrystalline spark plasma sintered Sr$_2$TiO$_4$ substrates was investigated to determine phase formation and preferred epitaxial orientation relationships ($ORs$) for isostructural Ruddlesden-Popper (RP) heteroepitaxy, further developing the high-throughput synthetic approach called Combinatorial Substrate Epitaxy (CSE). Both grazing incidence X-ray diffraction and electron backscatter diffraction (EBSD) patterns of the film and substrate were indexable as single-phase RP-structured compounds. The optimal growth temperature (between 650 $^{\circ}$C and 800 $^{\circ}$C) was found to be 750 $^{\circ}$C using the maximum value of the average image quality (IQ) of the backscattered diffraction patterns. Films grew in a grain-over-grain pattern such that each Ca$_2$MnO$_4$ grain had a single $OR$ with the Sr$_2$TiO$_4$  grain on which it grew. Three primary $ORs$ described 47 out of 49 grain pairs that covered nearly all of RP orientation space. The first $OR$, found for 20 of the 49, was the expected RP unit-cell over RP unit-cell $OR$, expressed as [100][001]$_{film}$||[100][001]$_{sub}$. The other two $ORs$ were essentially rotated from the first by 90$^{\circ}$, with one (observed for 17 of 49 pairs) being rotated about the [100] and the other  (observed for 10 of 49 pairs) being rotated about the [110] (and not exactly by 90$^{\circ}$). These results indicate that only a small number of $ORs$ are needed to describe isostructural RP heteroepitaxy and further demonstrate the potential of CSE in the design and growth of a wide range of complex functional oxides.
\end{abstract}


\maketitle

\newpage

\section{Introduction}





The Ruddlesden-Popper (RP) type phases are perovskite-like layered oxides of the general formula $A_{n+1}M_n$O$_{3n+1}$ (or $A$O($AM$O$_3$)$_n$),\cite{Ruddlesden:1957ky, Ruddlesden:1958dq, Sharma:1998uy} where $A$ is typically a rare earth or alkaline earth ion and $M$ is a transition or post-transition metal ion. The perovskite-like end-members of this family are the $n$ = $\infty$ $AM$O$_3$ perovskites and the $n$ = 1 $A_2M$O$_4$ RP phases (a structure also known a the K$_2$NiF$_4$ structure\cite{Ruddlesden:1957ky, Sharma:1998uy}). The structure of the RP family is described by the stacking of $n$ consecutive perovskite layers, $AM$O$_3$, alternating with a single $A$O rock-salt layer along the crystallographic $c$-direction. A schematic of the RP structure is shown in Figure 1 for the $n$ = 1 end-member. The reduced dimensionality of the $M$-O bonding plays an important role in the properties of RP phases, but the large unit cells and anisotropic nature of the RP crystal structures make synthesis of high-quality crystals, powders, and films more challenging than that of the related perovskites.



RP phases are known to exhibit a range of fascinating functional electronic properties, which can be tuned by changing the nature, size, and valence state of the $A$ and $M$ cations, non-stoichiometry and chemical order on cation and anion lattices, as well as through tailoring the dimensionality of the structure, i.e., the $n$ value.\cite{Sharma:1998uy, Dwivedi:1991vv, Raveau,Bednorz,Battle:1992vj, Attfield:1992ws, Lee:1993fr, Mahesh, Battle:1997ws, Greenblatt:1997fe, Fawcett:1998fk, Battle:1999dw, Kimura:2000fd, Lichtenberg:2002eb, Amow:2006ip, Cheong:2007fa, Nakhmanson:2008ko, Zhang:2013iy, Balachandran:2014jp,Mitchell1,Mitchell2,Mitchell3} For example, the layered cuprate RP phases played important roles in the discovery of high-temperature superconductivity,\cite{Sharma:1998uy, Dwivedi:1991vv,Raveau,Bednorz} and the manganese-based RP family display insulator-to-metal and ferromagnetic transformations at low temperatures, yielding colossal magnetoresistive properties.\cite{Sharma:1998uy, Mahesh, Fawcett:1998fk, Battle:1999dw, Kimura:2000fd,Mitchell1,Mitchell2,Mitchell3} The structural, transport and magnetic properties of nickelate RP compounds have also been extensively investigated,\cite{Greenblatt:1997fe, Kakol} being of recent interest as cathode materials in solid oxide fuel cells (SOFCs).\cite{Amow:2006ip} Other exotic properties include p-wave superconductivity in the RP ruthenate Sr$_2$RuO$_4$, \cite{Lichtenberg:2002eb} novel insulating electronic ground states from spin-orbit-coupling in Sr$_2$IrO$_4$,\cite{Zhang:2013iy, Miao:2014fj} and non-centrosymmetry in Ca$_2$IrO$_4$.\cite{Balachandran:2014jp} Even the first discovered RP family remains important, as Sr$_2$TiO$_4$ ($n$ = 1) presents advantageous properties for applications in tunable devices and as an alternative gate oxide in metal-oxide semiconductor field-effect transistors (MOSFETs),\cite{Haeni}  Sr$_4$Ti$_3$O$_{10}$ ($n$ = 3) has been reported for water-splitting photocatalytic activity without any co-catalyst,\cite{Ko} and novel ordered-structures in the titanate RPs are predicted to be interesting ferroelectrics.\cite{Nakhmanson:2008ko} The discovery of the Ca$_{n+1}$Mn$_n$O$_{3n+1}$ series followed quickly after the initial description of Sr$_{n+1}$Ti$_n$O$_{3n+1}$.\cite{Ruddlesden:1957ky, Ruddlesden:1958dq,RP2} These Mn-based materials are of interest for various electronic phenomena, such as insulator-to-metal transitions, charge-ordering, and colossal magnetoresistance. \cite{RPR}

RP phases are typically synthesized through conventional powder/ceramic synthesis. For practical electronic applications, to manipulate properties using substrate-induced strains, or to generate metastable compositions in the RP structure, thin films are of interest. Of course, the substrate choice is of utmost importance in designing and preparing high-quality films. Because of the good crystallographic relationship with the (001) RP plane and their widespread commercial availability, (100)-oriented perovskite single-crystals have been used as substrates for epitaxial growth of RP thin films ($n$ = 1 or 2 usually). However, this choice forces the growth direction to be the [001], which is kinetically more difficult to grow (i.e., requires stricter growth conditions to yield high quality films) and leads to out-of-phase boundaries owing to the existence of energetically degenerate nucleation events.\cite{Zurbuchen} Also, the growth of higher order members ($n$ > 2) usually requires advanced layer-by-layer deposition techniques, including reactive-molecular-beam-epitaxy (MBE) or Laser-MBE, and RHEED monitoring.\cite{Schlom,Matt,Matt2} These factors render the epitaxial RP growth more challenging than the simpler and more well-studied perovskite, and also restricts our understanding of growth largely to the [001] orientation.

For achieving the best epitaxy, one would ideally use an isostructural substrate, but the choice of commercial single-crystal RP substrates is extremely limited, including NdCaAlO$_4$ ($a$ = 3.685 \AA,  $c$ = 12.12 \AA), SrLaAlO$_4$ ($a$ = 3.76 \AA, $c$ = 12.63 \AA), and SrLaGaO$_4$ ($a$ =3.84 \AA,  $c$ = 12.68 \AA). These crystals have small $a$ lattice constants compared to many of the $A_2M$O$_4$ phases described above. Crystal growth on these substrates is often still carried out along the [001],\cite{locquet,weidong,bozo} but several studies have explored $a$-axis growth on the commercially-available (100) face, and these films generally grow with high quality, likely owing to the lower kinetic barriers to growth and fewer degenerate nucleation events.\cite{RPTF1,RPTF2,RPTF3} To understand better RP film growth and to exploit fully the potential of thin film design, it is of interest to broaden the types of $RP$ substrates on which one can grow. 

Rather than depositing on commercial single crystals, we have been developing a new strategy that uses well-characterized polycrystalline samples where each grain of the substrate can be viewed as a single crystal of a particular orientation.\cite{CSE1,CSE2,CSE3,Pravat,BFO}  We call this approach Combinatorial Substrate Epitaxy (CSE), in which a very large number of crystallographic orientations are present at the surface of the sample, and one can make statistical analyses of the growth modes across epitaxial orientation space. We have previously demonstrated with CSE, using typical growth conditions to obtain epitaxial films on single crystals, that: (1) grain-over-grain epitaxy occurs for most orientations of substrate grains, (2) only a small number of orientation relationships, $ORs$, exist for simple structures, (3) similar results are obtained for the growth of heterostructured complex oxide films using CSE as are obtained for films grown on single crystals, and (4) metastable and new complex oxides can be stabilized using CSE. The purpose of this work is to determine if similar observations can be made for RP phases using CSE on isostructural RP substrates, ultimately to improve our understanding of RP film growth.

Here, we use CSE to investigate the growth of the Ca$_2$MnO$_4$ ($a$ = 3.668 \AA, $c$ = 12.050 \AA) thin films on Sr$_2$TiO$_4$ ($a$ = 3.884 \AA, $c$ = 12.600 \AA).\cite{Ruddlesden:1957ky,RPR,RPparam}
 We focus on $n$ = 1 end-members, Ca$_2$MnO$_4$, of the RP oxides, since they are the simplest from a structural perspective. Indeed, layered perovskites are more difficult to synthesize in single-crystalline form than cubic perovskites, and they also exhibit various phenomena including high-T$_C$ or magnetoresistance. The Ca$_2$MnO$_4$ has not be investigated so far, and is thus of potential interest.
 Ca$_2$MnO$_4$ thin films were grown using standard pulsed laser deposition (PLD) similarly to previous works on La$_{2-x}$Sr$_x$NiO$_4$ or Sr$_{2-x}$La$_x$VO$_4$.\cite{RPTF3,RPTF4} 
We chose Sr$_2$TiO$_4$ as a substrate material because it is the archetypical $n$ = 1 RP oxide, but single crystals cannot be purchased. Instead, we prepared in-house spark plasma sintered polycrystalline substrates. Structural analyses were made using electron backscatter diffraction (EBSD) and grazing--incidence X-ray diffraction (GXRD). It will be shown that most of the film grew in a grain-over-grain fashion, regardless of the substrate surface orientation, and that most film grains adopted one of only three preferred epitaxial orientations.

\section{Experimental}


Commercial powders of SrCO$_3$ and TiO$_2$ (Cerac, with 99.5 \% and 99.9 \% purity, respectively) were used as precursors to synthesize polycrystalline Sr$_2$TiO$_4$ samples. The precursors were weighed in stoichiometric proportions, were attrition milled for $\sim$ 1h to reduce and homogenize the grain sizes, and then were calcined 1h at 1200 $^{\circ}$C. The calcined powders were ground in an agate mortar and then loaded between the two punches of a cylindrical graphite die, and protected from external pollution by graphite papers along the die walls. The die was then inserted into an SPS apparatus (Struers Tegra Force-5) to sinter the powder during 20 minutes at 1100 $^{\circ}$C, under a uniaxial load of 50 MPa. Using this process, larger densities can be obtained in shorter times than using conventional sintering routes.\cite{SPS1,SPS2} 

A highly dense pellet of 20 mm diameter was obtained and the structural quality of Sr$_2$TiO$_4$ was confirmed using X-ray diffraction (not shown). Substrates of $\sim$ 5 x 2 x 2 mm$^3$ were cut from the SPS pellets, where the surface of each substrate was taken as the face parallel to the applied pressure. These substrates were meticulously polished for EBSD characterization and film growth, which require flat surfaces of high crystalline quality (in EBSD only the first tenths of nanometers under the surface are probed\cite{Randle,BabaKishi:1998ur} and the nucleation events are sensitive to the near surface roughness and structure). First, several mechanical polishing steps were performed using SiC papers of different grain sizes, down to 10 $\mu$m, to obtain surfaces without any polishing marks. Second, the samples were polished automatically using diamond liquid pastes of grain sizes 3 $\mu$m and then 1 $\mu$m, to get mirror-like surfaces. Ca$_2$MnO$_4$ targets were sintered using classical solid state routes. 

Thin films (150 nm) were deposited by PLD on the surfaces of the as-prepared Sr$_2$TiO$_4$ substrates. Depositions were carried out using an O$_2$ pressure of 10$^{-3}$ mbar, a laser repetition rate of 2 Hz, and a target-to-substrate distance of 50 mm, leading to a deposition rate around 0.1 \AA /pulse. For any given deposition, the substrate temperature was fixed at a value between 650 to 800 $^{\circ}$C. The typical thickness of the films were about 30 nm. The composition was verified using energy-dispersive X-ray spectroscopy (EDS).

The substrates and the films were all analyzed by EBSD, using the commercial Orientation Imaging Microscopy (OIM$^{TM}$ from EDAX-AMETEK, Inc.) software (v.6.2). 
 This is a recently developed technique based on the acquisition of electron diffraction patterns from a sample placed in a scanning electron microscope (SEM).\cite{EBSD1} The sample is tilted to an angle of 70$^{\circ}$ from the horizontal to maximize the backscattering of electrons.\cite{EBSD2} EBSD patterns are formed during a two-stage diffraction process.\cite{EBSD3} Briefly, the incident electrons that enter the sample are first scattered inelastically. During this step the shape/dimensions of the interaction volume are determined. In the second stage of pattern formation, the inelastically scattered electrons travel towards the surface of the tilted sample and interact elastically with the crystal lattice planes. The result is that the bands observed in the diffraction patterns are the trace of a given family of planes, and the intersections of the different bands represent the relevant zone axes in the volume analyzed. As such, each Kikuchi (EBSD) pattern corresponds to a particular crystallographic orientation.\cite{EBSD1}

Using the crystallographic parameters of the phase, the software compares the experimental positions of the Kikuchi bands with the calculated positions of these bands.\cite{EBSD4} Several parameters are used to describe the EBSD data, including the Image Quality (IQ), which is a quantitative measure of the intensity of the Hough peaks and describes the crystal quality from the volume that contributes to a measured EBSD pattern,\cite{EBSD5} and the Confidence Index (CI), which measures the difference between the best and second best indexing solutions based on a voting algorithm (and ranges from 0 to 1). To compare the patterns of the film and the substrate, typical conditions were used: an SEM voltage of 20 kV, an aperture of 120 $\mu$m, and a working distance of 15 mm. By scanning the surface of the sample with a beam step size of 0.3 $\mu$m, using a hexagonal grid, many Kikuchi patterns are registered.\cite{EBSD4} The software assigns a color pixel for each orientation, and inverse pole figure (IPF) maps of the surface of the substrate and the film are thus recorded. To "clean" the data, points with a confidence index below 0.15 or misorientation angle greater than 2$^{\circ}$ from the average of the grain were removed, and colored black.

GXRD scans were collected with a grazing incident angle ranging from 0.3 to 1$^{\circ}$, to reduce the substrate contribution to X-ray patterns. A Bruker D8 Discover diffractometer was used in $\theta-\theta$ geometry with a Gobel mirror providing a parallel X-ray beam for a better control of the incident angle. A Cu anode with a K$\alpha_1$ line at 1.54056 {\AA} and a K$\alpha_2$ line at 1.54439 {\AA} was used.

 \section{Results and Discussion}
The EBSD patterns from both the films and substrates generally exhibited strong contrast and were well indexed by the automated software. Fig. 2 displays the average IQ measured over many film grains (of different orientations) deposited at a given temperature, for temperatures ranging from 650 to 800 $^{\circ}$C. Representative EBSD patterns from 30 nm thick films are given as insets for depositions carried out at 650, 700, and 750 $^{\circ}$C, which illustrate the evolution of the sharpness and brightness of bands. A maximum in the average IQ value occurs between 725 and 750 $^{\circ}$C (a maximum was also observed in the average CI for the films deposited at 750 $^{\circ}$C). Since these quantitative metrics of crystalline quality maximized for films deposited at 750 $^{\circ}$C, this temperature was used for the remainder of the Ca$_2$MnO$_4$ films studied in the present work. The increase in crystal quality with increasing temperature is likely associated with improved kinetics during nucleation. The decrease in the average IQ (and CI) at higher temperatures is unclear at this time, but may result from changes to a number of temperature dependent parameters, such as a decreased oxygen stoichiometry or increased roughness values, which in turn affect either crystal quality or EBSD image quality. 

It should be noted that similar optimizations of growth parameters are routinely made for films deposited on single-crystals, but using X-ray methods (e.g., the full-width-at-half-maximum in the $\omega$ - scan and/or the $\theta-2\theta$ scan, the shape of diffraction peaks, the absence of impurities, etc.) In CSE, conventional X-ray methods are not well-suited to the purpose of deposition parameter optimization, because many similarly located peaks arise from both the polycrystalline substrate and film. In the initial investigations of CSE-prepared films, choice of deposition parameters were mainly based on parameters optimized for similar films on single crystals.\cite{CSE1,CSE2,Pravat,BFO} These results provide a methodology to optimize growth parameters for CSE-prepared films. While we chose to optimize the average IQ as a function of temperature, one could do similar optimizations for any growth parameter, or even for a smaller subset of film orientations. In previous works on RP thin films deposited on single-crystals, the value of the full-width-at-half-maximum of the main diffraction peak was considered to optimize the structural quality.\cite{RPTF3,RPTF4} 


As done previously for CSE grown complex titanate films,\cite{CSE2} one can determine the epitaxial nature of growth by inspection of the EBSD pattern registered from the same grain before and after growth. Fig. 3 shows typical Kikuchi patterns recorded at the surface of the substrate and the film, for two different substrate grains (grain \# 5 and \# 9 discussed later in Fig. 5. The Kikuchi diagrams corresponding to the substrate (Fig. 3a and 3b) present sharper and brighter bands than those corresponding to the film measured at the same location (Fig. 3c and 3d), which is probably a result of relaxation of epitaxial mismatch. Nevertheless, the patterns from the film are of sufficient quality to be characterized using automated software. Furthermore, the bands and zone axes are in the same exact place in the two patterns, showing that the films grew in a unit-cell over unit-cell epitaxial fashion, in these locations. The quality of the films' patterns confirm the smooth flat surfaces, as expected from the average RMS roughness value around 1 nm.



To further confirm the formation of the RP phase, GXRD was performed.  Fig. 4 displays patterns recorded respectively at 1$^{\circ}$ from the surface for the substrate and 1, 0.50 and 0.30$^{\circ}$ for a 150 nm thick film. For comparison, a theoretical pattern for Ca$_2$MnO$_4$ was added. Using the material density, the depth probed by the X-ray beam is estimated to be of the order of the film thickness (150 nm) at 0.30$^{\circ}$ (within the experimental alignment errors). Overall, a dramatic increase of the films' signal-to-background ratio is observed as the angle is reduced. More precisely, small diffraction peaks appear close to 2$\theta$ = 33, 34.6, and 49.7$^{\circ}$ when the angle is reduced (a small contribution of Sr$_2$TiO$_4$ substrate is still observed even at 0.30$^{\circ}$). These reflections are assigned to the $103$, $110$, and $200$, respectively, of an $I4/mmm$ space group (as expected from an RP phase), and are well matched to the expected reflections of the Ca$_2$MnO$_4$ phase. The analysis therefore suggests the formation of an epilayer of the Ca$_2$MnO$_4$ phase on top of the Sr$_2$TiO$_4$ polycrystalline substrate. The corresponding lattice parameters of the film are calculated to be $a$ = 3.679 \AA, and $c$ = 12.24 \AA, which are slightly larger to the expected lattice parameters ($a$ = 3.668 \AA, $c$ = 12.050 \AA).\cite{Ruddlesden:1957ky,RPR,RPparam} 

 While confirming phase formation and providing average lattice parameters, the X-ray data do not yield local orientation relationships.


To understand the local growth and epitaxial relationships, IPF maps of the surface of the substrate and the film were recorded from the same area, and are shown in 5a and 5b. The IPF triangle color code is given in Fig. 5d. In these maps, points with poor or erroneous indexing (a CI < 0.15) have been colored black; such points correspond to less than 0.6 \% for the substrate scan and less than 6.8 \% for the film scan.  The average CIs for each map increased from 0.50 to 0.87 for the substrate and 0.18 to 0.67 for the film, respectively for before and after data cleaning. Enlarged images are shown in Fig. 5c for the substrate and Fig. 5e for the film. Several grains are clearly recognized, and are identified by numbers 1 to 9. For all of these grains (and most of those observed throughout this investigation), the films grew in a grain-over-grain fashion, where the film has a nearly uniform orientation over the entire substrate grain, terminating at the grain boundary. Moreover, the majority of grains display the same color (orientation) for the substrate and the film grains (\# 1, 2, 4, 5, 8 and 9), indicating the crystallographic orientations of these grains are similar to each other. In other words, these grains grew in a unit-cell over unit-cell fashion, even though they have a range of substrate orientations at the surface. Other grains (\# 3, 6 and 7) display completely different colors in the IPF maps, indicating that these film grains have a different orientation from the substrate grains, yet they still grew in a grain-over-grain fashion. 
 
The relationships between the substrate (red circles) and the film (black squares) are summarized in the Inverse Pole Figure (IPF) of Fig. 5f, for these 9 grains. For the grains \# 1, 2, 4, 5, 8 and 9, the symbols corresponding to the same grains on substrate and film are very close in the IPF, and the misorientation angles between substrate and film grains are less than 7$^{\circ}$. These values are indicative of nearly perfect unit-cell over unit-cell growth. As an example, grain \# 4 is growing with nearly a (100) orientation a similarly oriented substrate grain. The situation is, however,  more complex for grains \# 3, 6, and 7. Grain \# 3 has a (4,1,20) orientation for the substrate, but the same grain is growing near a (211) direction of the film.   This change in the epitaxial relationships can be described as a misalignment (tilt) in the grain direction with respect to that of the substrate, and will be discussed at length below. These observation indicate that the growth of these RP films on RP substrates are still influenced by grain orientation, but that only one orientation generally survives to the surface of the film. 

We sampled forty more grain pairs, to have a total of 49 observations of grain-over-grain epitaxial growth, to understand better the preferred epitaxial orientations during RP Ca$_2$MnO$_4$ on RP Sr$_2$TiO$_4$ film growth.  The $ORs$ between the film/substrate pairs were determined using software and methods described in Zhang et al.\cite{CSE1} To illustrate the small number of $ORs$, we plot in Fig. 6 the angle between the [100]$_{film}$ and [100]$_{sub}$ (on the horizontal axis) versus the angle between the [001]$_{film}$ and [001]$_{sub}$ (on the vertical axis). Of these 49 grains, only 4 distinct $ORs$ were observed, represented by clusters of points in the plot (and marked as such). Three primary $ORs$ described 47 out of 49 of the grain pairs, and we will focus on these three in our discussion. OR1 was found for 20 of the 49 pairs (including grains \# 1, 2, 4, 5, 8 and 9 discussed above), is represented by a cluster of points near the origin (lower left) in the figure, can be expressed as [100][001]$_{film}$||[100][001]$_{sub}$, and is described as the RP unit-cell over RP unit-cell $OR$. The slight offset from perfect alignment and the variation about the mean are related to relaxations of epitaxial mismatch, as observed elsewhere in CSE grown films.\cite{CSE1, CSE3} The inset marked OR1 indicates the basic crystallographic alignment of the two RP cells for this $OR$. This may be considered the trivial $OR$ case, as the two unit cells align with one another and, therefore, should have a low-energy (coherent) interface, regardless of the orientation. That this relationship is not observed everywhere is more surprising than the fact that this $OR$ is the most frequently observed $OR$, at about 40 percent of the population. When RP phases are grown on RP single crystals, this is the most frequently discussed $OR$.\cite{Zurbuchen} 


OR2 was found for 17 of the 49 pairs (including grain \# 3 discussed above), is represented by a cluster of points near the upper left corner of the figure, can be expressed as [100][001]$_{film}$||[100][010]$_{sub}$, and is described as being rotated from OR1 by 90$^{\circ}$ about the [100]. The inset marked OR2 indicates the basic crystallographic alignment of the two RP cells for this $OR$ (not accounting for the slight rotations related to epitaxial mismatch). It should be noted that common defects in [001]-oriented RP films (grown on (001) RP or perovskite single crystal surfaces) are misoriented regions where the c-axis is rotated by 90$^{\circ}$ about the <100> axes of the RP structure. \cite{Zurbuchen} 
 While this defect structure is often considered to arise as a result of unfavorable kinetic conditions, which do not allow for the c-axis to align out of plane, its widespread observation in thin films is an indication that such interfacial planes are of relatively low energy. Therefore, it is not surprising to find OR2 as the second most frequently observed $OR$, and just slightly less frequent than OR1. Presumably, this orientation is similarly related to kinetic growth factors that inhibit perfect growth of OR1, while also providing a relatively low energy interface. More work is required to understand whether this OR initiates at the substrate interface directly, or occurs during growth, and the preferentially coarsens until the whole grain has a singular $OR$. At this thickness and in these specific growth conditions, slightly less about 35 percent of grains exhibit this $OR$ over the entire grain. Overall, about three quarters of all grains exhibited OR1 or OR2. 

OR3 was found for 10 of the 49 pairs (including grain \# 3 and 6 discussed above), is represented by a cluster of points near the upper central region of the figure, and is slightly more complicated to describe than the other $ORs$. If the cluster of points had been located exactly at 45$^{\circ}$ on the x axis and 90$^{\circ}$ on the y axis, then this $OR$ could be expressed as [001][110]$_{film}$||[110][001]$_{sub}$ and could described as being rotated from OR1 by 90$^{\circ}$ about the [110]. The inset marked OR3 indicates the basic crystallographic alignment of the two RP cells for this approximate $OR$. However, the cluster of points is not quite at 90$^{\circ}$ on the y-axis. The actual alignment is about 16$^{\circ}$ away from this 90$^{\circ}$ rotation, and can be expressed as [223][991]$_{film}$||[110][001]$_{sub}$. The angle between the [223] and [991] is about 89$^{\circ}$, so they can align reasonably well with the 90$^{\circ}$ difference between the [110] and [001]. The difference between the approximate relationship, shown as the inset OR3, and the real $OR$ is simply a 16$^{\circ}$ rotation about the [110]. The cause of this "extra" rotation is unclear at this time, and represents an interesting observation made possible by CSE. To our knowledge, this $OR$ represents a new $OR$ for RP film growth and about 20 percent of all grains have this $OR$ over the entire grain. As for OR2, more work is required to understand whether OR3 initiates at the substrate interface directly, or occurs during growth, and the preferentially coarsens until the whole grain has a singular $OR$. Overall, about 95 percent all grains exhibited OR1, OR2, or OR3. 
 
An obvious question to ask is: are the different $ORs$ described above correlated in some way to the substrate surface normal of the grain on which it grew? Rather than plotting the pairs of substrates on the same stereographic triangle, as done in Fig. 5f, we plot the four ORs using different symbols on two stereographic triangles in Fig. 7. In Fig. 7a, the $ORs$ are plotted with respect to the substrate surface normal, while in Fig. 7b they are plotted with respect to the film surface normal. Largely speaking, OR1 and OR2 are distributed throughout the stereogram, with a slight increase in OR2 in the central region of the triangle. OR3 shows up almost exclusively near the edge of the triangle, near [110] extending in the direction of [001]. For single crystal growth on RP single crystals, the [110] orientations have not been investigated, and this may explain why OR3 has not been reported. OR4 was observed only near the [001] oriented substrate crystals. With only two observations, it is difficult to make any strong conclusion here (the epitaxy of OR4 can be expressed as [212][520]$_{film}$||[110][001]$_{sub}$), and nearby both OR1 and OR2 are also observed. 
 

It is also interesting to explore how the ORs are distributed with respect to the film surface normal, as shown in Fig. 7b. As expected OR1 remains distributed throughout the triangle (with decreased frequencies where OR2 and OR3 had clusters in Fig. 7b. OR2 is also distributed throughout, but has a larger cluster of observation near the [110] orientation (related to the cluster in the center of Fig. 6(b), which have rotations similar to grain 7 in Fig. 5f. Some of OR1 and OR2 are clustered with an orientation near [001], as are almost all of OR3. What is surprising about this observation is that growth along the [001] is considered kinetically challenging owing to the large repeat period along that direction. For OR3, where the c-axis is nearly in the plane of the substrate surface, the film grows with the c-axis nearly perpendicular to the substrate surface. This observation seems counterintuitive to the accepted kinetic models of RP film growth, where c-axis films are difficult to obtain, and implies that there is more to understand for epitaxial growth on general surfaces of crystals.Perhaps, c-axis films could be obtained more readily on RP orientations near the {110} since the latest being the most thermodynamically stable planes. More work is necessary to unravel the details and origin of this seemingly anomalous OR, as is true also for the nature of OR2. 
 
\section{Conclusions}


Ruddlesden-Popper (RP) type Ca$_2$MnO$_4$ thin films were grown by pulsed laser deposition on isostructural Sr$_2$TiO$_4$ polycrystalline substrates prepared by spark plasma sintering (SPS). Optimization of the film structural quality as a function of deposition temperature was made by considering the pattern quality from electron backscatter diffraction. Grazing-incidence X-ray diffraction of the film deposited at 750 $^{\circ}$C confirmed the RP phase formation. The Kikuchi patterns displayed bright and sharp bands, attesting to the local formation of the RP phases everywhere on the substrate. Inverse Pole Figure maps recorded on the same grain before and after film deposition revealed a perfect grain-over-grain growth for the majority of grains. Three primary $ORs$ described 47 out of 49 grain pairs substrate-film that covered nearly all of RP orientation space. The first $OR$, found for 20 of the 49, was the expected RP unit-cell over RP unit-cell $OR$, expressed as [100][001]$_{film}$||[100][001]$_{sub}$. The other two $ORs$ were essentially rotated from the first by 90$^{\circ}$, with one (observed for 17 of 49 pairs) being rotated about the [100] and the other  (observed for 10 of 49 pairs) being rotated about the [110] (and not exactly by 90$^{\circ}$). These results indicate that only a small number of $ORs$ are needed to describe isostructural RP heteroepitaxy and further demonstrate the potential of CSE in the design and growth of a wide range of complex functional oxides.

We thank L. Gouleuf and J. Lecourt for technical support. M. Lacotte received her PhD scholarship from the Ministere de l'Enseignement Superieur et de la Recherche. D. Pravarthana is supported by a PhD fellowship included in the Erasmus Mundus Project IDS-FunMat. Partial support of the French Agence Nationale de la Recherche (ANR), through the program Investissements d'Avenir (ANR-10-LABEX-09-01), LabEx EMC3 and the Interreg IVA MEET project are also acknowledged. We also thank Dr. O Perez, Dr. I. Canero-Infante, Prof. B. Mercey and B. Raveau for fruitful discussions.


\newpage
Figures Captions:

Figure 1 : Schematic of the tetragonal $A_2M$O$_4$ Ruddlesden-Popper (or $K_2Ni$F$_4$) structure. The perspective view is looking along the $a$-axis, with the $b$-axis ($c$-axis) along the horizontal (vertical). The spheres represent the $A$ cations. The $M$ cations (oxygen) sit at the center (vertices) of the octahedra, which share corners in the $a$-$b$ plane but are not connected along the $c$-axis.

Figure 2:  Graphical representation of the growth optimization methodology used for Ca$_2$MnO$_4$ films deposited on Sr$_2$TiO$_4$ polycrystalline substrates. The average image quality (IQ) measured over many grains is plotted as a function of the deposition temperature. Example EBSD (Kikuchi) patterns are also given, and the automated determination of crystallographic orientations are given in Euler angles ($\varphi$$_1$,$\Phi$,$\varphi$$_2$). Units are degrees.

Figure 3 : Kikuchi patterns recorded from two substrate grains, (a) and (b), and two film grains, (c) and (d), grown upon them. (a) and (c) were registered from grain \# 5 and (b) and (d) from grain \# 9 (discussed later in Fig. 5). The relevant zone axes are indicated in white rectangles. The respective orientations corresponding to (a), (b), (c) and (d) are, in Euler angles, (217,80,297), (231,83,254), (219,84,295), and (234,87,253). Units are degrees.

Figure 4: XRD patterns recorded respectively at 1$^{\circ}$ from the surface for the Sr$_2$TiO$_4$ substrate and 1, 0.50 and 0.30$^{\circ}$ for a Ca$_2$MnO$_4$ film of thickness 150 nm grown upon a Sr$_2$TiO$_4$ substrate. The theoretical pattern of the Ca$_2$MnO$_4$ phase is shown for comparison.

Figure 5 : Inverse pole--figure (orientation) maps of the surface of the substrate (a),(c) and the film (b),(e), where (c) and (e) are enlargements of the indicated areas of (a) and (b). The color code for these maps is given as the standard stereographic triangle of a tetragonal crystal in (d). Orientations of the 9 grains in (c) and (e) are plotted in (f) on the the standard stereographic triangle, where the numbers correspond to the grain number and a red circle or black square indicate respectively a substrate or film grain. The small areas with confidence indices less than the acceptable threshold are colored black in (a-e).

Figure 6 : A plot of the angles of misorientation between the [100] (horizontal axis) and [001] (vertical axis) for the film and substrate for the 49 grain pairs plotted in Fig. 7. The insets provide the approximate relationship between the unit cells for $ORs$ 1, 2, and 3. Note that $OR$ 3 is depicted as a 90$^{\circ}$ rotation about [110], which is an oversimplified description (see text). Axes for film and substrate are in black and blue, respectively.

Figure 7 : Orientations of 49 grain pairs that were used to determine the epitaxial $ORs$, plotted using the standard stereographic triangle. In both (a) and (b), the marker type indicates what $OR$ was found for the grain with that orientation, as indicated in the key. (a) Marker locations indicate the orientation of the substrate, Sr$_2$TiO$_4$. (b) Marker locations indicate the orientation of the film, Ca$_2$MnO$_4$.


\begin{thebibliography}{99}
\bibitem{Ruddlesden:1957ky} S. Ruddlesden, P. Popper, Acta Crystallogr. 10 (1957) 538. 
\bibitem{Ruddlesden:1958dq} S. Ruddlesden, P. Popper, Acta Crystallogr. 11 (1958) 54.
\bibitem{Sharma:1998uy} I. Sharma, D. Singh, Bull. Mater. Sci. 21 (1998) 363.
\bibitem{Dwivedi:1991vv} A. Dwivedi and A. N. Cormack, Bulletin Of Materials Science, 14 (1991) 575.
\bibitem{Raveau}  N. Nguyen, J. Choisnet, M. Hervieu, B. Raveau, J. Solid State Chem. 39 (1981) 120. 
\bibitem{Bednorz} J.G. Bednorz and K.A. Muller, Z. Phys. B 64 (1986) 189.
\bibitem{Battle:1992vj} P.D. Battle, S.K. Bollen, A.V. Powell, J. Solid State Chem. 99 (1992) 267.
\bibitem{Attfield:1992ws} M.P. Attfield, P.D. Battle, S.K. Bollen, S.H. Kim, A.V. Powell, M. Workman, J. Solid State Chem. 96 (1992) 344.
\bibitem{Lee:1993fr} J.Y. Lee, J.S. Swinnea, H. Steinfink, W.M. Reiff, S. Pei, J.D. Jorgensen, J. Solid State Chem. 103 (1993) 1.
\bibitem{Mahesh} R. Mahesh, R. Mahendiran, A.K. Raychaudhuri, C.N.R. Rao, J. Solid State Chem. 122 (1996) 448.
\bibitem{Battle:1997ws} P.D. Battle, M.A. Green, N.S. Laskey, J.E. Millburn, M.J.R.L. Murphy, S.P. Sullivan, and J.F. Vente, Chem. Mater. 9, (1997) 552.
\bibitem{Greenblatt:1997fe} M. Greenblatt, Current Opinion in Solid State {\&} Materials Science, 2 (997) 174.
\bibitem{Fawcett:1998fk} I. D. Fawcett, Sunstrom, M. Greenblatt, M. Croft, and K. V. Ramanujachary, Chem. Mater. 10 (1998)  3643.
\bibitem{Battle:1999dw} P. D. Battle and M. J. Rosseinsky, Current Opinion in Solid State {\&} Materials Science, 4 (1999) 163.
\bibitem{Kimura:2000fd} T. Kimura and Y. Tokura, Annual Review of Materials Science, 30 (2000) 451.
\bibitem{Lichtenberg:2002eb} F. Lichtenberg, Progress in Solid State Chem., 30 (2002) 103.
\bibitem{Amow:2006ip} G. Amow and S. J. Skinner, J Solid State Electrochem, 10 (2006) 538.
\bibitem{Cheong:2007fa} S.-W. Cheong, Nature Mater. 6 (2007) 927.
\bibitem{Nakhmanson:2008ko} S. Nakhmanson, Phys. Rev. B 78 (2008) 064107.
\bibitem{Zhang:2013iy} H. Zhang, K. Haule, and D. Vanderbilt, Phys. Rev. Lett. 111 (2013) 246402.
\bibitem{Balachandran:2014jp} P. V. Balachandran, D. Puggioni, and J. M. Rondinelli, Inorg. Chem., 53 (2014) 336.
\bibitem{Mitchell1} C.D. Ling, J.E Millburn, J.F. Mitchell, D.N. Argyriou, J.  Linton, JH.N. Bordallo, Phys. Rev. B 62 (2000) 15096.
\bibitem{Mitchell2} D.N. Argyriou, J.F. Mitchell, P.G. Radaelli, H.N. Bordallo, D.E. Cox, M. Medarde, J.D Jorgensen, Phys. Rev. B 59 (1999) 8695.
\bibitem{Mitchell3} J.F. Mitchell, D.N. Argyriou, J.D. Jorgensen, D.G. Hinks, C.D. Potter, S.D. Bader, Phys. Rev. B 55 (1997) 63.
\bibitem{Kakol} Z. Kakol, J. Spalek, J.M. Honig, J. Solid State Chem. 79 (1989) 288.
\bibitem{Miao:2014fj} L. Miao, H. Xu, and Z. Q. Mao, Physical Review B, 89 (2014) 035109.
\bibitem{Haeni} J.H. Haeni, C.D. Theis, D.G. Schlom, W. Tian, X.Q. Pan, H. Chang, I. Takeuchi, X.-D. Xiang, Appl. Phys. Lett. 78 (2001) 3292.
\bibitem{Ko} Y.-G. Ko, W.Y. Lee, Catal. Lett. 83 (2002) 157.
\bibitem{RP2} M. Velazquez, C. Haut, B. Hennion, A. Revcolevshi, J. Cryst. Growth 220 (2000) 480.
\bibitem{RPR} C. Autret, C. Martin, M. Hervieu, R. Retoux, B. Raveau, G. Andre and F. Boure, J. Solid State Chem. 177 (2004) 2044.
\bibitem{Zurbuchen} M.A. Zurbuchen, W. Tian, X.Q. Pan, D. Fong, S.K. Streiffer, M.E. Hawley, J. Lettieri, Y. Jia, G. Asayama, S.J. Fulk, D.J. Comstock, S. Knapp, A.H. Carim, D.G. Schlom, J. Mater. Res. 22 (2007) 1439.
\bibitem{Schlom} W. Tian, J.H. Haeni, D.G. Schlom, E. Hutchinson, B.L. Sheu, M.M. Rosario, P. Schiffer, Y. Liu, M.A. Zurbuchen, X.Q. Pan, Appl. Phys. Lett. 90 (2007) 022507.
\bibitem{Matt} R.G. Palgrave, P. Borisov, M.S. Dyer, S.R.C. McMitchell, G.R. Darling, J.B. Claridge, M. Batuk, H. Tan, H. Tian, J. Verbeeck, J. Hadermann, M.J. Rosseinsky, J. Am. Chem. Soc. 134 (2012) 7700.
\bibitem{Matt2} Lei Yan, Hongjun Niu, Craig, A. Bridges, P.A. Marshall, J. Hadermann, G. van?Tendeloo, P.R. Chalker and M.J. Rosseinsky, Angew. Chem. 46 (2007) 4539.
\bibitem{locquet} J.-P. Locquet, J. Perret, J. Fompeyrine, E. Machler, J.W. Seo, G. Van Tendeloo, Nature 394 (1998) 453.
\bibitem{weidong} Weidong Si and X. X. Xi, Appl. Phys. Lett. 78 (2001) 240.
\bibitem{bozo} I. Bozovic, G. Logvenov, I. Belca, B. Narimbetov, and I. Sveklo,  Phys. Rev. Lett. 89 (2002) 107001.
\bibitem{RPTF1} S. Madhavan, D.G. Schlom, A. Dabkowski, H.A. Dabkowska, Y. Liu, Appl. Phys. Lett. 68 (1996) 559.  
\bibitem{RPTF2} J. Matsuno, Y. Okimoto, M. Kawasaki and Y. Tokura, Phys. Rev. Lett. 95 (2005) 176404. 
\bibitem{RPTF3} S. Shinomori, M. Kawasaki and Y. Tokura, Appl. Phys. Lett. 80 (2002) 574.
\bibitem{CSE1} Y. Zhang, A.M. Schultz, L. Li, H. Chien, P.A. Salvador, G.S. Rohrer, Acta Mater. 60 (2012) 6486.
\bibitem{CSE2} S. Havelia, S. Wang, K.R. Balasubramaniam, A.M. Schultz, G.S. Rohrer, P.A. Salvador, CrystEngComm 15 (2013) 5434.
\bibitem{CSE3} A.M. Schultz, Y. Zhu, S.A. Bojarski, G.S. Rohrer, P.A. Salvador, Thin Solid Films 548 (2013) 220.
\bibitem{Pravat} D. Pravarthana, O.I. Lebedev, S. Hebert, D. Chateigner, P. A. Salvador and W. Prellier, Appl. Phys. Lett. 103 (2013) 143123.
\bibitem{BFO} D. Pravarthana, M. Trassin, Jiun Haw Chu, M. Lacotte, A. David, R. Ramesh, P. A. Salvador, and W. Prellier, Appl. Phys. Lett. 104 (2014) 082914.
\bibitem{RPparam} M.E. Leonowicz, K.R. Poeppelmeier and J.M. Longo, J. Solid State Chem. 59, 71-80 (1985) 
\bibitem{RPTF4} J. Matsuno, Y. Okimoto, M. Kawasaki and Y. Tokura, Appl. Phys. Lett. 82 (2003) 194.
\bibitem{SPS1} Z.A. Munir, D.V. Quach, M. Ohyanagi, J. Am. Ceram. Soc. 94 (2011) 1.
\bibitem{SPS2} M. Nygren, Z. Shen, Solid State Sciences 5 (2003) 125.
\bibitem{Randle} V. Randle, Materials Characterization 60 (2009) 913.
\bibitem{BabaKishi:1998ur} K.Z. Baba-Kishi, Scanning 20 (1998) 117.
\bibitem{EBSD1} A.J. Schwartz, M. Kumar, B.L. Adams, D.P. Field, Electron Backscatter Diffraction in Materials Science, Springer (2009).
\bibitem{EBSD2} A. Winkelmann, G. Nolze, Ultramicros. 110 (2010) 190.
\bibitem{EBSD3} R.P. Goehner, J.R. Michael, J. Res. Natl. Inst. Stand. Technol. 101 (1996) 301.
\bibitem{EBSD4} B.L. Adams, S.I. Wright, K. Kunze, Metallurgical Transactions A 24 (1993) 819.
\bibitem{EBSD5} S.I Wright and M.M. Nowell, Microscopy Microanalysis, 12 (2006) 72. 

 \end{thebibliography}
\end{document}